\documentclass{pasj00}

\begin{document}
\SetRunningHead{T. Nagayama et al.}{NH$_3$ emission in CMZ}
\Received{2006/11/25}

\title{A Complete Survey of the Central Molecular Zone in NH$_3$}

\author{Takumi \textsc{Nagayama},\altaffilmark{1}
        Toshihiro \textsc{Omodaka},\altaffilmark{2}
        Toshihiro \textsc{Handa},\altaffilmark{3} \\
        Hayati Bebe Hajra \textsc{Iahak},\altaffilmark{1}
        Tsuyoshi \textsc{Sawada},\altaffilmark{4}
        Takeshi \textsc{Miyaji},\altaffilmark{5} \\
        and Yasuhiro \textsc{Koyama}\altaffilmark{6}}
\altaffiltext{1}{Graduate School of Science and Engineering, 
		 Kagoshima University, \\
                 1-21-30 K$\hat{o}$rimoto, Kagoshima 890-0065}
\altaffiltext{2}{Faculty of Science, Kagoshima University,
                 1-21-30 K$\hat{o}$rimoto, Kagoshima 890-0065}
\altaffiltext{3}{Institute of Astronomy, University of Tokyo, 
                 2-21-1 Osawa, Mitaka, Tokyo 181-0015}
\altaffiltext{4}{Nobeyama Radio Observatory, National Astronomical 
                 Observatory of Japan, \\
                 Minamimaki, Minamisaku, Nagano 384-1305}
\altaffiltext{5}{Mizusawa VERA Observatory, 
	 	 National Astronomical Observatory of Japan, \\
		 2-21-1 Osawa, Mitaka, Tokyo 181-8588} 
\altaffiltext{6}{Kashima Space Research Center, National Institute of 
                 Information and Communications Technology,\\
                 893-1 Hirai, Kashima, Ibaraki 314-8510} 
\email{nagayama@astro.sci.kagoshima-u.ac.jp}


\maketitle

\begin{abstract}
We present a map of the major part of the central molecular zone (CMZ)
of simultaneous observations in the NH$_3$ $(J,K)$ = (1,1) and (2,2) lines 
using the Kagoshima 6-m telescope. 
The mapped area is 
$\timeform{-1D.000} \leq l \leq \timeform{1D.625}$, 
$\timeform{-0D.375} \leq b \leq \timeform{+0D.250}$. 
The kinetic temperatures derived from the (2,2) to (1,1) intensity 
ratios are 20--80 K or exceed 80 K.
The gases corresponding to temperature of 20--80 K and $\geq$ 80 K 
contain 75\% and 25\% of the total NH$_3$ flux, respectively.
These temperatures indicate that 
the dense molecular gas in the CMZ is dominated by gas
that is warmer than the majority of the dust present there.
A comparison with the CO survey by \citet{sawada} shows that 
the NH$_3$ emitting region is surrounded by 
a high pressure region on the $l$-$v$ plane.
Although NH$_3$ emission traces dense gas, 
it is not extended into a high pressure region.
Therefore, 
the high pressure region is less dense and has to be hotter.
This indicates that 
the molecular cloud complex in the Galactic center region has 
a ``core'' of dense and warm clouds which are traced by the NH$_3$ emission, 
and an ``envelope'' of less dense and hotter gas clouds.
Besides heating by ambipolar diffusion, 
the hot plasma gas emitting the X-ray emission
may heat the hot ``envelope''.

\end{abstract}


\section{Introduction}

Gas temperature is one of the basic parameters 
that control star formation activity. 
In the same volume, 
the interstellar medium is believed to be isothermal and 
the gas and dust temperatures should be the same. 
However, 
several surveys of the interstellar medium in the Galactic center 
region reveal different temperatures of gas and dust.

The dust temperature is measured in the submillimeter/infrared continuum.
It is found to be as cold as 20 K in the 450 $\mu$m and 850 $\mu$m continuum 
by the Submillimetre Common-User Bolometer Array (\cite{pierce}) 
or 15--22 K in the 45--175 $\mu$m continuum 
by the Infrared Space Observatory (\cite{lis}).

However, 
the molecular gas is warmer than the dust.
In the millimeter line observations made using the IRAM 30-m telescope, 
the gas temperature is found to be as high as 60--70 K by multi-line analysis 
(\cite{lis}).
A more direct estimation of the gas temperature 
is derived from NH$_3$ line observations.
\citet{morris} observed 
the NH$_3$ $(J,K)$ = (1,1), (2,2), and (3,3) lines of 
the Galactic center region at $b = -\timeform{2'}$ 
using the NRAO 11-m telescope.
They concluded that 
the kinetic temperature is uniformly between 30 and 60 K.
\citet{hutt} observed 
the NH$_3$ $(J,K)$ = (1,1), (2,2) and (4,4), (5,5) lines 
of selected clouds using the NRAO 43-m telescope 
and a few small maps using the Bonn 100-m telescope.
They reported two temperature components of gas; 
the first, 
derived from the intensity ratio of (2,2) to (1,1), 
was as low as 25 K, 
while the second, 
derived from the intensity ratio of (5,5) to (4,4),
was as high as 200 K.

These NH$_3$ surveys suggest that 
the gas in the Galactic center would be a mixture of hot and cold gases. 
However, 
the previous surveys cover only limited regions, 
such as on a single strip along the Galactic plane or only selected clouds.
Therefore, 
a large-scale survey is required to comprehensively study the conditions 
of the entire molecular gas in the central molecular zone (CMZ).

We have observed the CMZ in the NH$_3$ lines to investigate the dense gas.
We present the large-scale data of the Galactic center in the NH$_3$ 
$(J,K)$ = (1,1) and (2,2) lines for the first time .
In section 2, 
we describe the observations.
The data are presented in section 3.
In section 4, 
we discuss the physical conditions of the molecular gas in the CMZ 
through comparisons with previous observations.
In this paper, 
we assume that the distance to the Galactic center is 8.5 kpc.
For the direction and position in the sky, 
we use Galactic east as the positive Galactic longitude, 
and Galactic north as the positive Galactic latitude.


\section{Observations}

\subsection{Data from the Kagoshima 6-m Telescope}

We conducted a large-scale survey using the Kagoshima 6-m telescope of the
National Astronomical Observatory of Japan (NAOJ) from September 2000 to 
April 2002.
We made simultaneous observations in two inversion transitions of 
the NH$_3$ $(J,K)$ = (1,1) and (2,2) lines at 23.694495 and 23.722633 GHz, 
respectively.
At a wavelength of 1.3 cm, the telescope beamwidth is \timeform{9'.5} 
and the main beam efficiency ($\eta_{\mathrm{mb}}$) is 0.59.
We used a $K$-band HEMT amplifier whose system noise temperature is 
200--300 K and a 2048-channel TeO$_2$ crystal acoustic-optical 
spectrometer with a bandwidth of 250 MHz and frequency resolution of 
250 kHz.
At the NH$_3$ (1,1) and (2,2) frequencies, these correspond to 
3200 km s$^{-1}$ velocity coverage and 3.2 km s$^{-1}$ velocity 
resolution.
We obtained approximately 250 NH$_3$ (1,1) and (2,2) spectra at
$\timeform{-1D.000} \leq l \leq \timeform{1D.625}$ and
$\timeform{-0D.375} \leq b \leq \timeform{+0D.250}$
with a spacing of \timeform{0D.125}.
The surveyed area corresponds to 390$\times$90 pc based on 
the distance to the Galactic center, 8.5 kpc.
All data were obtained by position switching 
between the target positions and reference positions.
The reference positions were obtained at the Galactic latitude 
$b < \timeform{-1D}$, 
where neither NH$_3$ (1,1) nor (2,2) emission was detected.
We integrated at least 30 min at each point.
The relative pointing error is better than \timeform{1'}, which was 
verified by the observations of several H$_2$O (frequency 22.235080 GHz) 
maser sources.

Data reduction was performed using the UltraSTAR
package developed by the radio astronomy group at the University of Tokyo
(\cite{nakajima}, in preparation).
To improve the signal to noise ratio, 
the obtained spectra are smoothed to a velocity resolution of 5 
or 10 km s$^{-1}$.
The rms noise level after 5 km s$^{-1}$ smoothing is typically 0.080 K
in the unit of the main beam brightness temperature defined by  
$T_{\mathrm{mb}} \equiv T_{\mathrm{A}}^{*} / \eta_{\mathrm{mb}}$, 
where $T_{\mathrm{A}}^{*}$ is the antenna temperature 
calibrated by the chopper wheel method (\cite{kutner}).
In this paper, the intensities are presented in the main beam 
temperature.


\subsection{Data from the Kashima 34-m Telescope}

In order to confirm the ``\timeform{0D.9} wing feature'' 
(see section 3.6),
we carried out a single point observation at 
$(l,~b) = (\timeform{0D.880},~\timeform{0D.000})$.
This observation was made using the \timeform{1'.6} beam of 
the Kashima 34-m telescope
of the National institute of Information and Communications Technology 
(NICT) on April 18, 2006.
The single point spectra in the NH$_3$ (1,1) and (2,2) lines 
were obtained by the same method using the Kagoshima 6-m telescope survey.


\section{Results}

\subsection{Profiles}

\begin{figure*}[h]
  \begin{center}
    \FigureFile(80mm,30mm){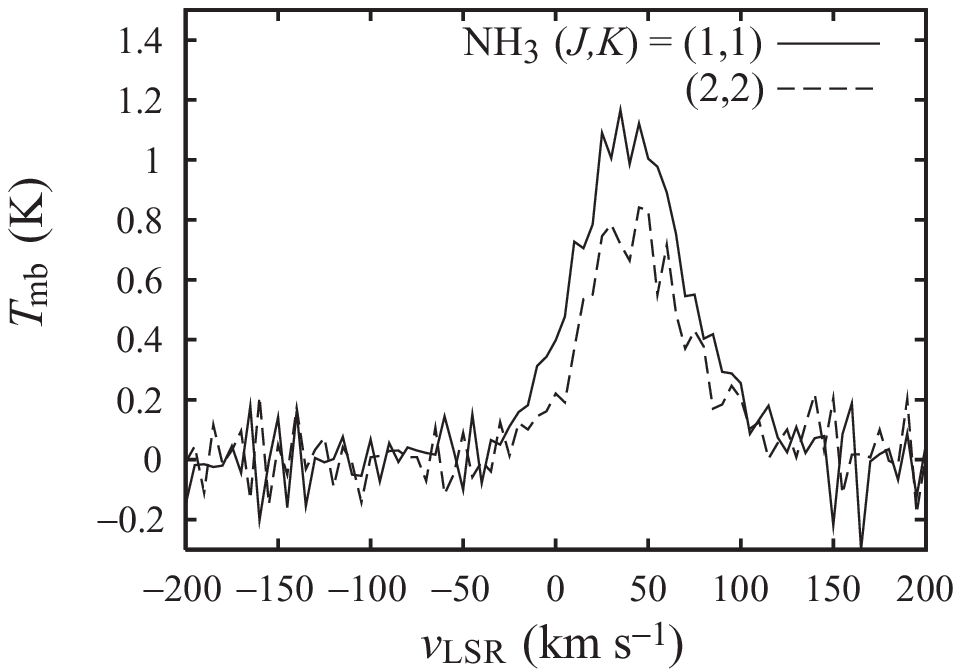}
  \end{center}
  \caption{Spectra of NH$_3$ (1,1) (solid line) and (2,2) (dashed line)
           at $(l,~b) = (\timeform{0D.750},~\timeform{-0D.125})$.
           Both spectra are at a 5 km s$^{-1}$
           velocity resolution and are shown in the $T_{\mathrm{mb}}$ unit.}
  \label{fig:profile}
\end{figure*}

Figure \ref{fig:profile} shows the line profiles of NH$_3$ (1,1) and (2,2) 
at $(l,~b) = (\timeform{0D.750},~\timeform{-0D.125})$
near the giant molecular cloud Sgr B shown 
with a velocity resolution of 5 km s$^{-1}$. 
The line shapes of the two NH$_3$ transitions are very similar over 
a 150 km s$^{-1}$ range.

In quiescent clouds, the NH$_3$ line profile comprises five 
quadruple hyperfine components consisting of a main line and two 
symmetric pairs of satellite lines.
The intensity ratio of the main line and the satellite lines provides 
the optical depth of each rotational level, 
and a unique rotational temperature is 
derived from the ratio of the optical depth at different rotational levels.

In the Galactic center region, 
molecular emission is extended in the sky
and shows violent motion even in the comparatively small area
covered by a single telescope beam.
In fact, 
none of our data profiles show satellite lines above the noise level.
Therefore, it is difficult to separate the five quadruple hyperfine 
components from the observed spectra with a large beam.
However, 
we can derive the rotational and kinetic temperatures 
without the opacity estimation (see section 4.1).


\subsection{Integrated Intensity Distribution}

\begin{figure*}[h]
  \begin{center}
    \FigureFile(160mm,20mm){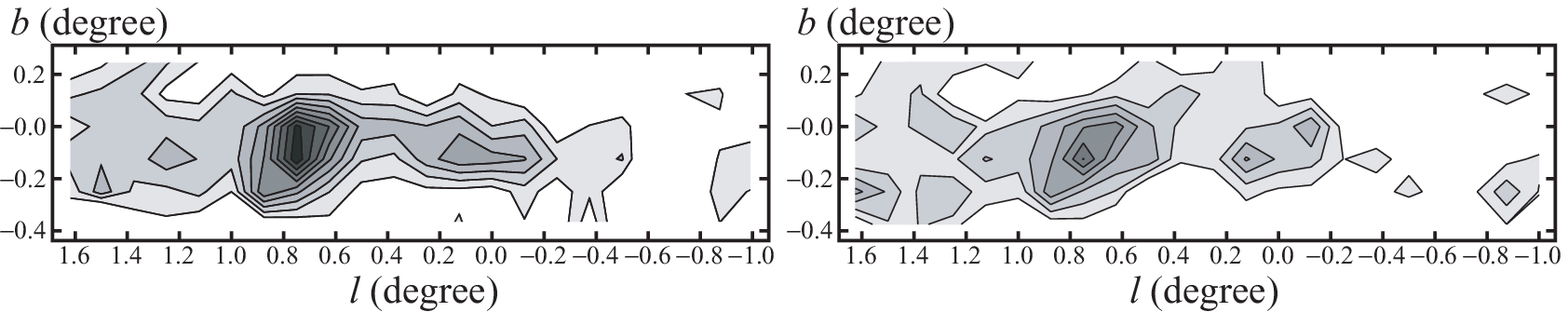}
  \end{center}
  \caption{Integrated intensity maps of NH$_3$ (1,1) (left) and (2,2)
           (right).
           The lowest contour and the contour interval are
           8.5 K km s$^{-1}$ in $\int T_{\mathrm{mb}} dv$.
           Velocity range of integration is
           $-200 \leq v_{\mathrm{LSR}} \leq 200$ km s$^{-1}$.}
  \label{fig:integ}
\end{figure*}

Figure \ref{fig:integ} shows the integrated intensity maps of NH$_3$ (1,1) 
and (2,2).
Our entire observed area almost covers the CMZ.
The overall distribution is well traced in other molecular emission lines
such as the CO (\cite{sawada}) and CS (\cite{tsuboi}) lines. 
The total integrated intensity in the entire observed area of NH$_3$ 
(1,1) and (2,2) is
$\int T_{\mathrm{mb}}(1,1) dv = 1580$ K km s$^{-1}$ and 
$\int T_{\mathrm{mb}}(2,2) dv = 1330$ K km s$^{-1}$, respectively.
These values correspond to 
$2.26 \times 10^{5}$ Jy km s$^{-1}$ and 
$1.90 \times 10^{5}$ Jy km s$^{-1}$ in flux units.
The NH$_3$ emission is concentrated at
\timeform{-0D.500} $\leq l \leq$ \timeform{1D.625} 
and its distribution is asymmetrical with respect to Sgr A* at 
$l = \timeform{-0D.06}$.
The intensity of the NH$_3$ line located at the Galactic eastern side of 
Sgr A* is 81\% in the entire observed area, 
although it is 73\% in the CS line derived from a map by \citet{tsuboi}.
The NH$_3$ emissions are also extended along Galactic latitude.
The total intensity at the observed positions on $b = \timeform{0D}$,
effectively covering a strip of one beamwidth, 
is only 27\% in the entire observed area.
The overall distribution along Galactic latitude appears to be 
symmetric about $b = \timeform{-0D.05}$, where Sgr A* is located.
The full width at half maximum (FWHM) along Galactic latitude is 
$\Delta b =$ \timeform{0D.2}--\timeform{0D.3},
which is almost the same as CO; 
however, it is more confined at the \timeform{1D.3} region and at 
$l = \timeform{-0D.125}$.

The strongest NH$_3$ emission in our observations originates from 
the Sgr B cloud, 
which has a large number of H\emissiontype{II} regions.
The second prominent feature is seen near Sgr A corresponding to the 
Sgr A clouds. 
Another NH$_3$ cloud at $l = \timeform{1D.2}$ is often called the
\timeform{1D.3} region.
We summarize the integrated intensities for major regions in Table 
\ref{tab:flux}.


\subsection{Prominent Clouds}

We identify four NH$_3$ clouds
which both (1,1) and (2,2) intensities exceed the 3 $\sigma$ level.
All of them are identified to be the known molecular clouds seen in other 
molecular lines.

\begin{figure*}[h]
  \begin{center}
    \FigureFile(160mm,80mm){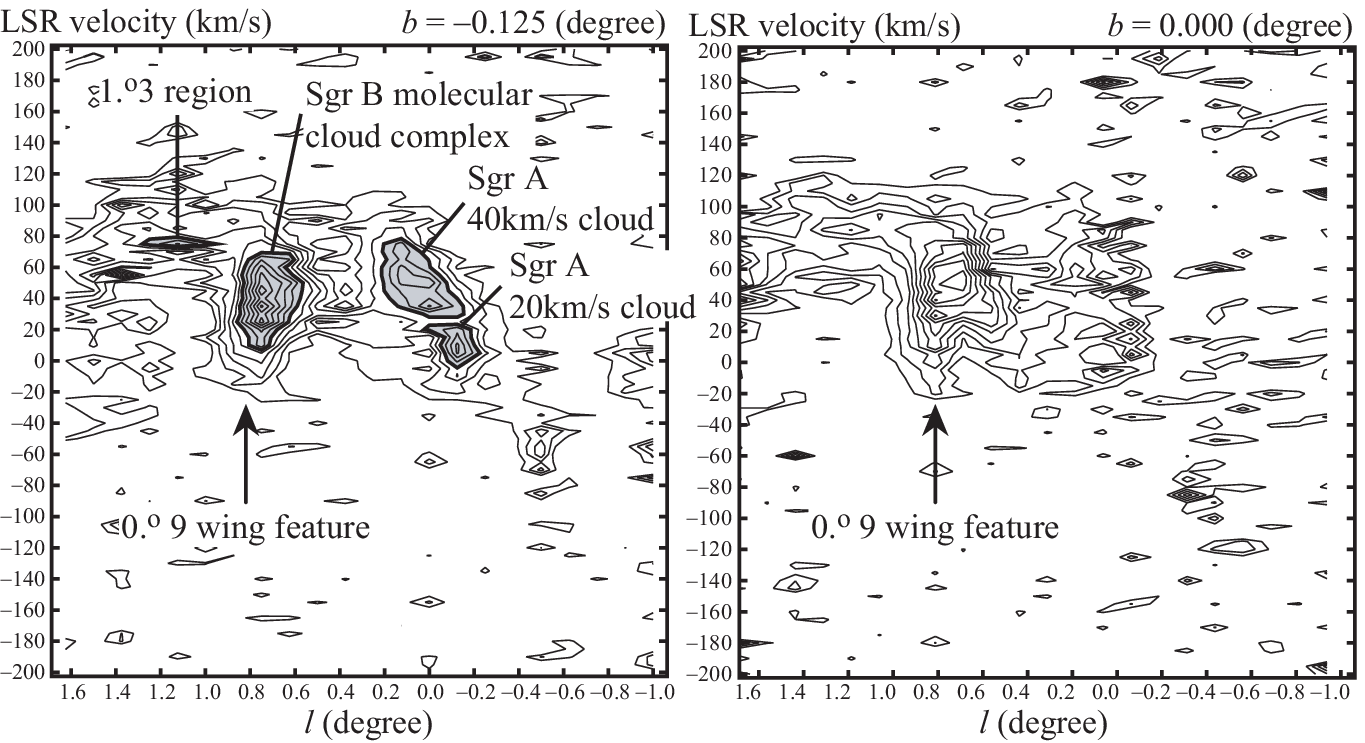}
  \end{center}
  \caption{The distribution of the identified clouds
          in the $l$-$v$ diagram of NH$_3$ (1,1) at
          $b = \timeform{-0D.125}$.}
  \label{fig:schematic}
\end{figure*}

The two clouds that appear at 
$(l,~b,~v_{\mathrm{LSR}}) \simeq 
(\timeform{-0D.125},~\timeform{-0D.125},~10~{\mathrm{km~s}}^{-1})$ and
$(l,~b,~v_{\mathrm{LSR}}) \simeq 
(\timeform{0D.125},~\timeform{-0D.125},~55~{\mathrm{km~s}}^{-1})$
are the ``Sgr A 20 km s$^{-1}$ cloud'' and ``Sgr A 40 km s$^{-1}$ cloud'', 
respectively, which were identified by \citet{gusten}.
A cloud appearing at 
$(l,~b,~v_{\mathrm{LSR}}) \simeq 
(\timeform{0D.750},~\timeform{-0D.125},~40~{\mathrm{km~s}}^{-1})$
is associated with the Sgr B1 and Sgr B2 H\emissiontype{II} regions.
Sgr B2 is one of the most massive star forming regions in our Galaxy.
A cloud appeared at 
$(l,~b,~v_{\mathrm{LSR}}) \simeq 
(\timeform{1D.125},~\timeform{-0D.125},~80~{\mathrm{km~s}}^{-1})$ 
is the \timeform{1D.3} region cloud. 
It is the weakest among the four identified clouds.
Although the \timeform{1D.3} region has a large latitudinal scale height 
($\Delta b \geqq \timeform{0D.5}$) in the CO emission (\cite{oka1}),
The NH$_3$ emitting area is as small as $\Delta b=$ \timeform{0D.2}.
The NH$_3$ emission is more confined than the CO emission in the 
\timeform{1D.3} region.
These four identified clouds in the $l$-$v$ space 
are shown in Figure \ref{fig:schematic}.


\subsection{``\timeform{0D.9} Wing Feature'': 
A High Velocity Wing at $l = \timeform{0D.9}$}

\begin{figure*}[h]
  \begin{center}
    \FigureFile(80mm,30mm){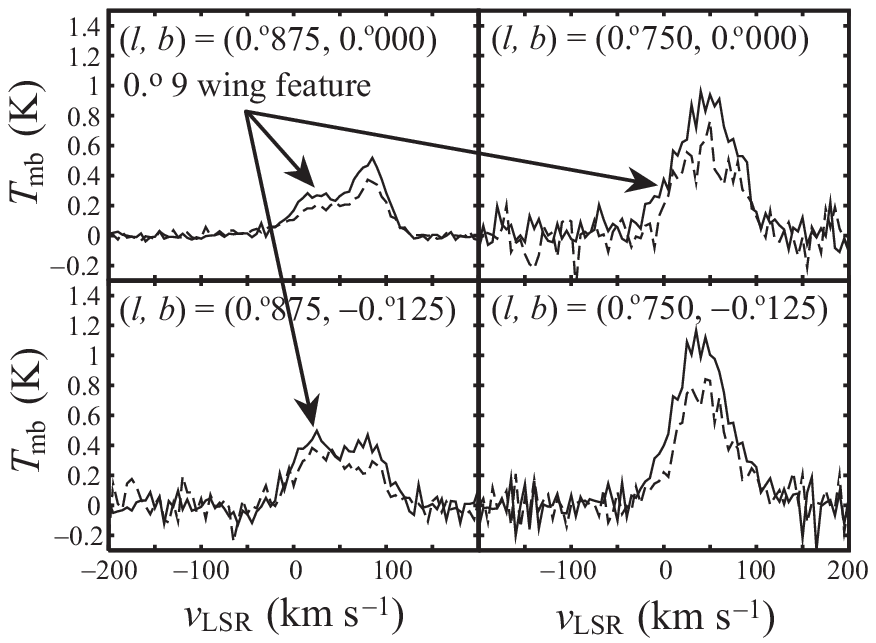}
  \end{center}
  \caption{Spectra of NH$_3$ (1,1) (solid line) and (2,2) (dashed line)
           around the ``\timeform{0D.9} wing feature''.
           These profiles are shown at 5 km s$^{-1}$ velocity resolution.
           The double peak components
           ($v_{\mathrm{LSR}} =$ 20 and 80 km s$^{-1}$) are seen at
           $(l,~b) = (\timeform{0D.875},~\timeform{0D.000})$ and
           (\timeform{0D.875},~\timeform{-0D.125}).
           The blueshifted wing is seen at
           $(l,~b) = (\timeform{0D.750},~\timeform{0D.000})$.}
  \label{fig:profile2}
\end{figure*}

In the $l$-$v$ diagrams at $b = \timeform{-0D.125}$ and \timeform{0D.000},
a prominent blueshifted wing is seen near $l = \timeform{0D.9}$.
This is shown in Figure \ref{fig:schematic} and \ref{fig:profile2}.
This wing is extended even with the \timeform{9'.5} beam.
We detected this wing feature at
\timeform{0D.750}  $\leq l \leq$ \timeform{0D.875} and
\timeform{-0D.125} $\leq b \leq$ \timeform{0D.000}.
The wing at $l = \timeform{0D.750}$ extends from
$v_{\mathrm{LSR}} = 0$ km s$^{-1}$ to $-30$ km s$^{-1}$.
It is blueshifted from the main spectral feature by 80 km s$^{-1}$. 
This blueshifted wing is also seen in previous observations of both the 
NH$_3$ (\cite{morris}) and CS (\cite{tsuboi}) emission,
although no one mentioned about this feature.
The wing is associated with the double peak profiles 
($v_{\mathrm{LSR}} = 20$ and 80 km s$^{-1}$)
at $(l,~b) = (\timeform{0D.875},~\timeform{0D.000}),~
(\timeform{0D.875},~\timeform{-0D.125})$.
Although the 80 km s$^{-1}$ component is the main ridge through the entire
Galactic center region linked with $l >$ \timeform{1D.000},
the 20 km s$^{-1}$ component is seen only at these four positions.
The FWHM of the 20 km s$^{-1}$ component is larger than 50 km s$^{-1}$.
The blueshifted wing of the 20 km s$^{-1}$ component is 
conspicuous in the $l$-$v$ diagram at $b = \timeform{0D.000}$,
whereas that in the $l$-$v$ diagram at $b = \timeform{-0D.125}$ 
is blended with the intense emission from Sgr B cloud.
However, the 20 km s$^{-1}$ component of 
$(l,~b) = (\timeform{0D.875},~\timeform{-0D.125})$
is stronger than that of $(l,~b) = (\timeform{0D.875},~\timeform{0D.000})$.

\begin{figure*}[h]
  \begin{center}
    \FigureFile(80mm,30mm){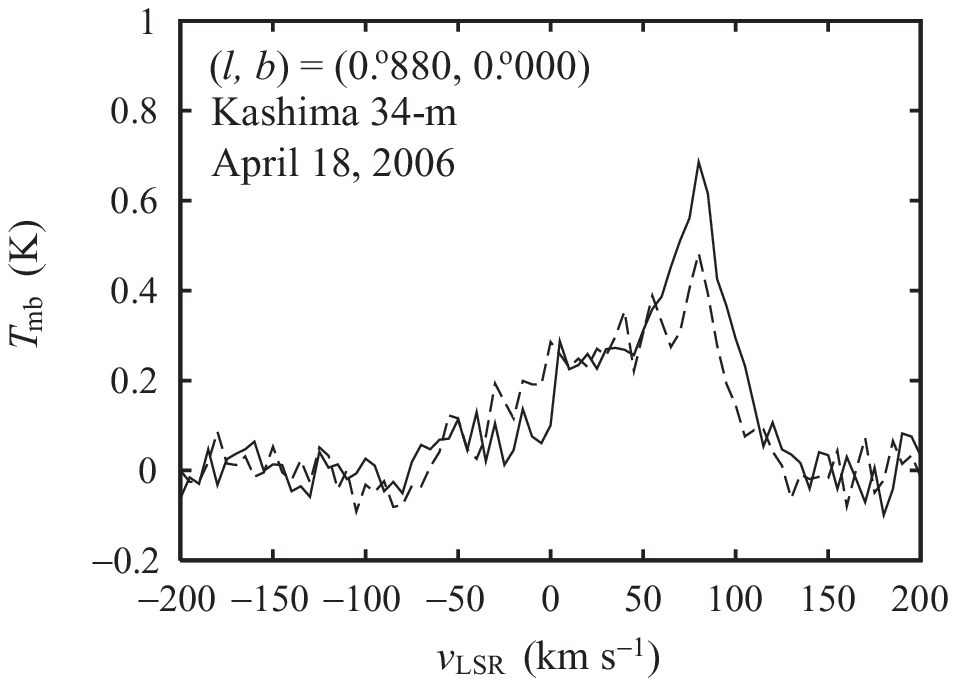}
  \end{center}
  \caption{Spectra of NH$_3$ (1,1) (solid line) and (2,2) (dashed line)
           at $(l,~b) = (\timeform{0D.880},~\timeform{0D.000}$)
           with 5 km s$^{-1}$ velocity resolution.
           These spectra are obtained using the Kashima 34-m telescope
           on April 18, 2006.}
  \label{fig:profile3}
\end{figure*}

We made higher resolution observations 
at $(l,~b) = (\timeform{0D.880},~\timeform{0D.000)})$ using the Kashima 
34-m telescope.
Figure \ref{fig:profile3} shows the spectra obtained in the (1,1) and (2,2) 
lines.
From these observations,
the blueshifted wing and the 20 km s$^{-1}$ component are confirmed in 
both the (1,1) and (2,2) lines.


\section{Discussion}

\subsection{Gas Temperature of the CMZ}

The intensity ratio of the (2,2) line to (1,1) line, $R_{(2,2)/(1,1)}$,
is controlled by the kinetic temperature and optical depth of NH$_3$ gas.
Previous estimations of the optical depth of NH$_3$ are 
$\tau \sim 3\mbox{--}10$ for the Sgr A 20 km s$^{-1}$ cloud (\cite{gusten}),
$\tau \sim 4$ in the stronger NH$_3$ emitting sources, and
$\tau \sim 2.3 \pm 1.0$ in the weaker sources (\cite{hutt}).
Because the flux of our observations originates from weaker and 
more extended sources than the previous observations,
the optical depth may be significantly smaller than these results.

The CS line can also trace dense gas.
In the Galactic center region, 
the CS line emission has moderate optical depth, 
$\tau \leq$ 2--3.
The total molecular mass in the CMZ is 
$M(\mathrm{H}_2) = (3\mbox{--}8) \times 10^{7} \MO$ (\cite{tsuboi}).
\citet{dahmen} estimate that the large scale 
$^{12}$CO ($J = 2\mbox{--}1$) emission in the CMZ
is of maderate ($\tau \gtrsim 1$) or low optical depth ($\tau < 1$)
and the total molecular mass in the CMZ is 
$M(\mathrm{H}_2) = (2\mbox{--}5) \times 10^{7} \MO$.
The total fluxes in the NH$_3$ (1,1) and (2,2) lines are 
$\int S dv = 2.26 \times 10^{5}$ and 
$1.90 \times 10^{5}$ Jy km s$^{-1}$, respectively.
In the case where the NH$_3$ emission is optically thin,
the total mass of the molecular gas in the whole region is
$M_{\mathrm{lum}} = 1 \times 10^{8} \MO$ based on 
$X(\mathrm{NH}_3) = 10^{-9}$ (\cite{hutt}).
The total masses derived from NH$_3$, CS, and CO are consistent.
This suggests that the NH$_3$ line should have moderate optical depth
or low optical depth.

Therefore, we estimate the rotational temperature $T_{\mathrm{rot}}$ 
from $R_{(2,2)/(1,1)}$ in the optically thin ($\tau \ll 1$) and 
optically thick ($\tau \sim 10$) cases using the method of \citet{morris}.
The conversion from $T_{\mathrm{rot}}$ to $T_{\mathrm{k}}$ is performed 
according to \citet{hutt}.
It shows that $T_{\mathrm{k}}$ is always higher than $T_{\mathrm{rot}}$.
When $T_{\mathrm{rot}}$ is less than 20 K, 
it is very close to $T_{\mathrm{k}}$, 
and when $T_{\mathrm{rot}}$ is 40 K, 
$T_{\mathrm{k}}$ is 80 K, which is almost two times higher than
$T_{\mathrm{rot}}$ (see Table \ref{tab:2}).

\begin{figure*}[h]
  \begin{center}
    \FigureFile(80mm,60mm){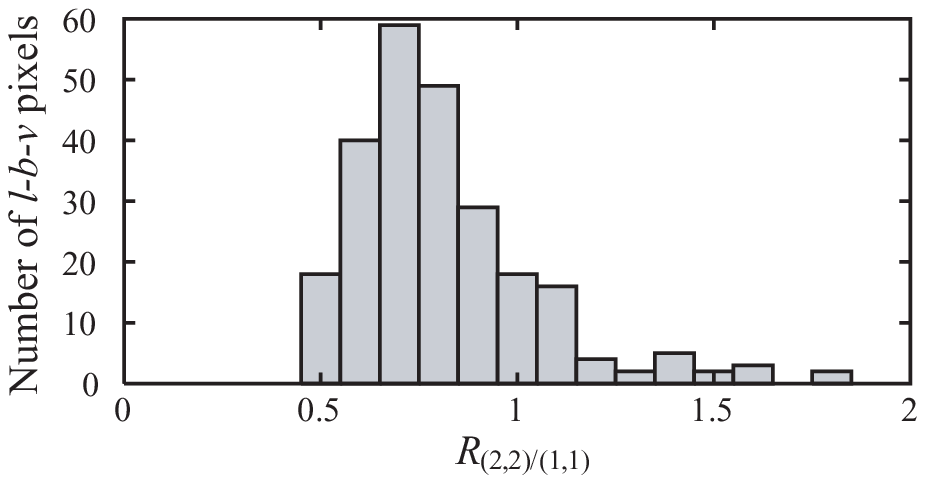}
  \end{center}
  \caption{The histogram of the intensity ratios of
           NH$_{3}$ (2,2) to (1,1).
           We count the $l$-$b$-$v$ pixels at
           which the line was detected over
           the 1.5 $\sigma$ level after 10 km s$^{-1}$ smoothing.}
  \label{fig:histogram}
\end{figure*}

Before we calculate the temperature, 
we examine the $R_{(2,2)/(1,1)}$ statistics.
Figure \ref{fig:histogram} shows a histogram of $R_{(2,2)/(1,1)}$ for 
pixels in the entire observed area at which an NH$_3$ (1,1) line 
was detected over a 1.5 $\sigma$ level after smoothing to 10 km s$^{-1}$.
The $R_{(2,2)/(1,1)}$ is within a range from 0.5 to 1.8 in our data.

Previous observations indicate the discrepancies in the gas and dust 
temperatures in the Galactic center.
The gas temperature is always higher than the dust temperature.
The hot gas temperature is approximately 100 K and 
the cold dust temperature is approximately 20 K.
For the obtained value of $R_{(2,2)/(1,1)}$,
we divide the ratios into three regimes 
($R_{(2,2)/(1,1)} =$ 0.5--0.6, $R_{(2,2)/(1,1)} =$ 0.7--0.8, and 
$R_{(2,2)/(1,1)} >$ 0.9)
which correspond to gas temperatures ranging from that of
cold dust (20 K) to hot gas (exceeding 100 K).

$R_{(2,2)/(1,1)} =$ 0.5--0.6 corresponds to 
$T_{\mathrm{rot}} \simeq$ 20--30 K and $T_{\mathrm{k}} \simeq$ 20--40 K.
The NH$_3$ flux with $R_{(2,2)/(1,1)} =$ 0.5--0.6 is approximately 25\% 
of the total NH$_3$ flux.
The low value of $T_{\mathrm{rot}}$ and $T_{\mathrm{k}}$
is close to the dust temperatures given by \citet{pierce} and \citet{lis}.
This low $R_{(2,2)/(1,1)}$ gas is as cool as the cold dust.

The gas with $R_{(2,2)/(1,1)} =$ 0.7--0.8, 
which corresponds to $T_{\mathrm{rot}} \simeq$ 30--40 K and 
$T_{\mathrm{k}} \simeq$ 40--80 K, is warmer than the cold dust.
The NH$_3$ flux of this warm gas is approximately 50\% of the total NH$_3$ flux.
$R_{(2,2)/(1,1)}$ of this dominant gas component is consistent with the 
ratios found by \citet{morris}.

\begin{figure*}[h]
  \begin{center}
    \FigureFile(160mm,160mm){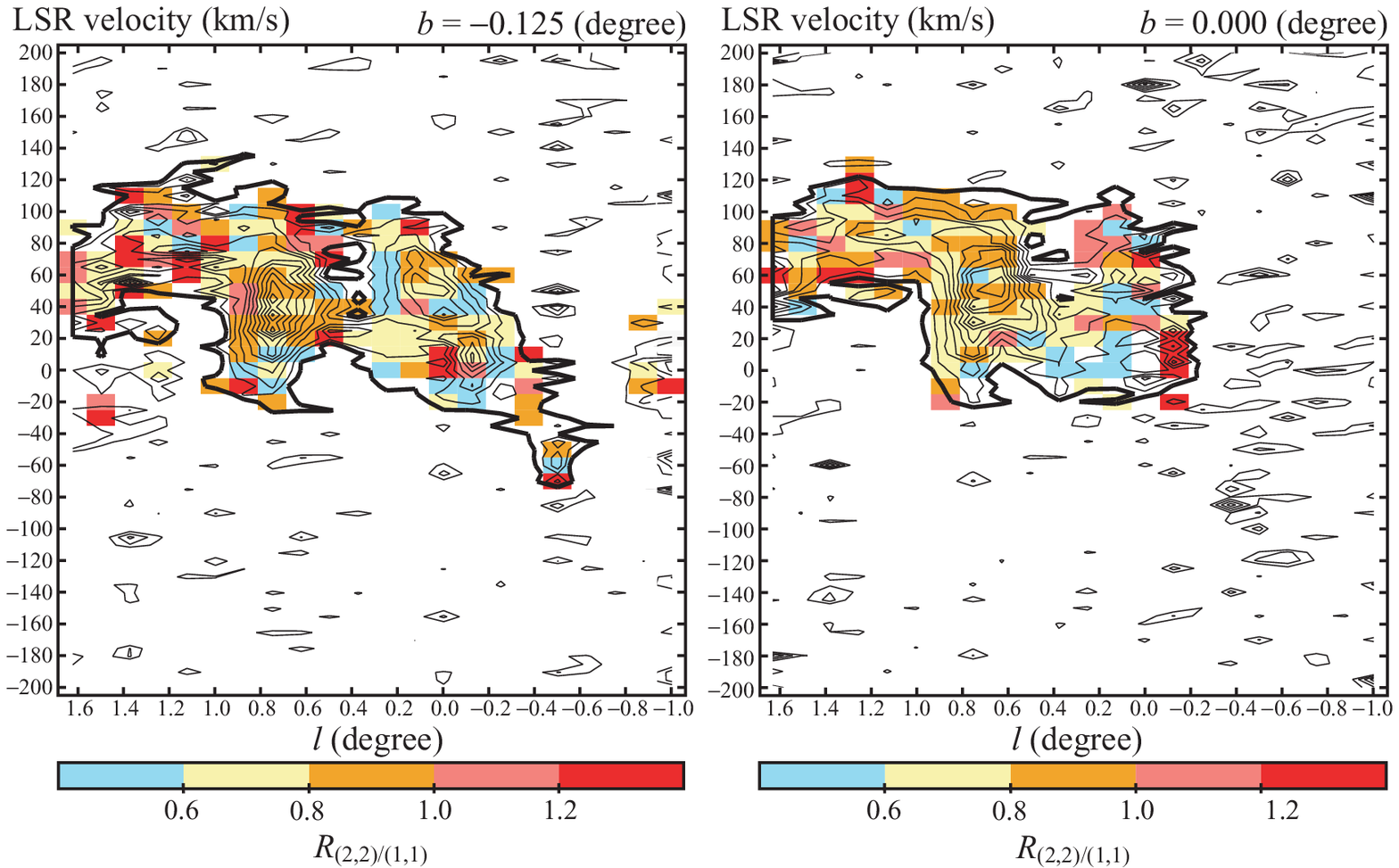}
    \caption{$l$-$v$ diagrams of the intensity ratios of NH$_3$ (2,2) to (1,1).
             The thick line shows the NH$_3$ emitting region
             surrounded by the NH$_3$ (1,1) lowest contour.}
    \label{fig:ratio}
  \end{center}
\end{figure*}

\begin{figure*}[h]
  \begin{center}
    \FigureFile(160mm,80mm){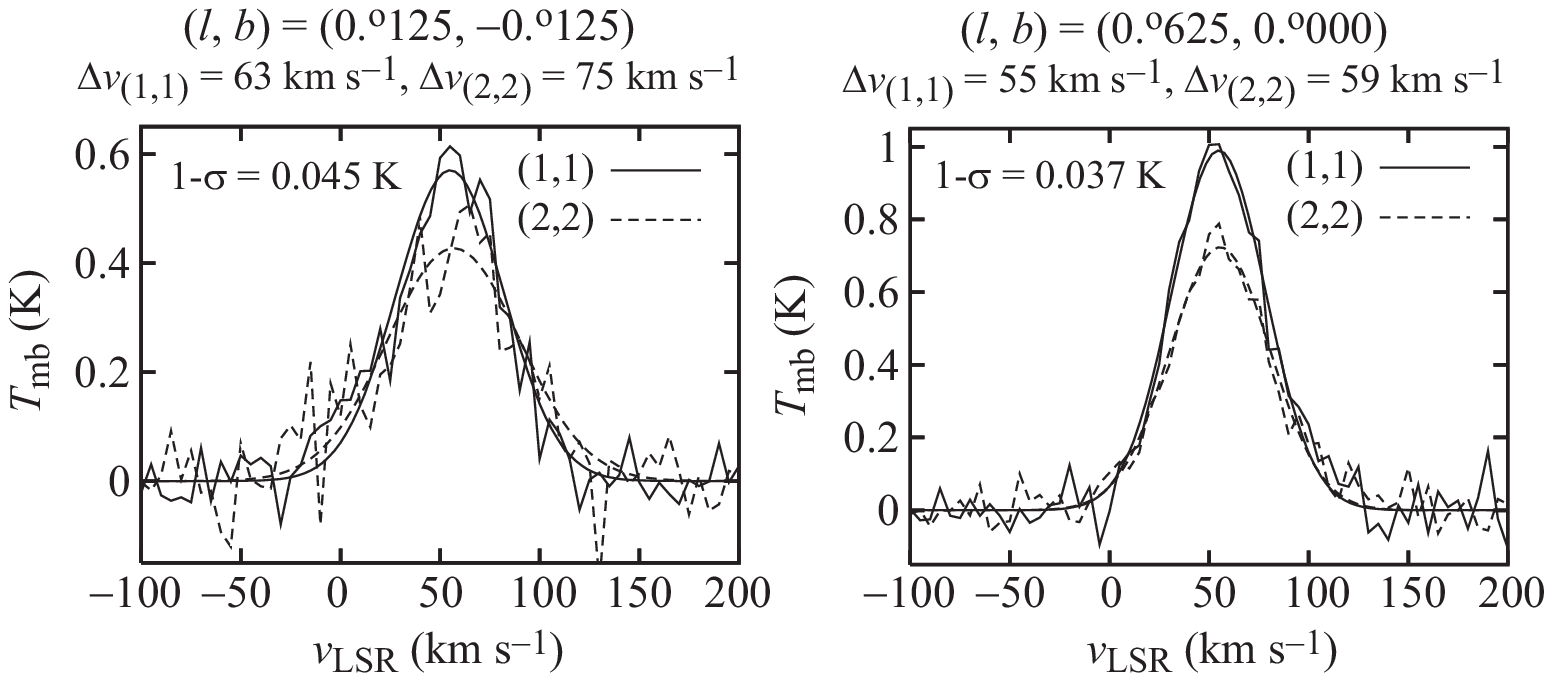}
  \end{center}
  \caption{Spectra of NH$_3$ (1,1) (solid line) and (2,2) (dashed line)
           at $(l,~b) = (\timeform{0D.125},~\timeform{-0D.125}$)
           and $(l,~b) = (\timeform{0D.625},~\timeform{0D.000}$)
           with 5 km s$^{-1}$ velocity resolution.
           The (2,2) spectra are broadened more than
           the (1,1) spectra.}
  \label{fig:sample}
\end{figure*}

The histogram also shows a small number of high ratio 
($R_{(2,2)/(1,1)} > 0.9$) pixels.
The NH$_3$ flux with this high $R_{(2,2)/(1,1)}$ is approximately 25\% 
of the total NH$_3$ flux.
These high $R_{(2,2)/(1,1)}$ pixels
appear at the edge of the NH$_3$ emitting region on the $l$-$v$ plane 
(see Figure \ref{fig:ratio}).
In fact, when we compare the obtained (1,1) and (2,2) profiles,
$R_{(2,2)/(1,1)}$ sometimes appears to increase in both the line edges.
Figure \ref{fig:sample} shows the spectra of NH$_3$ (1,1) and (2,2) 
at $(l,~b) = (\timeform{0D.125},~\timeform{-0D.125})$ and
$(l,~b) = (\timeform{0D.625},~\timeform{0D.000})$
with a velocity resolution of 5 km s$^{-1}$. 
These spectra show the higher (2,2)/(1,1) ratio at the edges of the profiles,
because the line width of (2,2) has extended more than that of (1,1).

Before considering this ratio enhancement due to gas property,
we should check effects of the satellite lines.
The satellite lines of (2,2) that have a wider frequency spacing 
than those of (1,1).
The separations of satellite lines in the (2,2) transition are 
$+26.02, +16.38, -16.38,$ and $-26.03$ km s$^{-1}$.
Those in the (1,1) transition are 
$+19.85, +7.46, -7.38,$ and $-19.56$ km s$^{-1}$.
This means that the broading due to the satellites in the (2,2)
transition might be more sufficient than that in the (1,1) transition.
To evaluate this effect, 
we examine it with the cases of some optical depths.
A typical NH$_3$ line exhibits a Gaussian shape profile
with a 60 km s$^{-1}$ width and the typical ratio of the NH$_3$ (2,2) line 
to the (1,1) line is 0.75.
In the case of moderate and low optical depth ($\tau < 1$), 
the satellite lines of (2,2) 
are much weaker than the main line and could not affect the observed line 
width of (2,2).
In the case of the high optical depth ($\tau > 30$),
the satellite lines affect the observed line width.
The wider separation in the (2,2) line results in greater broadening
than that in the case of the (1,1) line, which gives ratio enhancement 
at the line edge.
Thus, in this case, the higher ratio at the line edges were not real.
However, the NH$_3$ emission in the CMZ is not high optically thick 
($\tau \ll 30$).
Therefore, the higher ratio at the line edges 
means that the gas near the velocity edge is hotter than 
bulk of the cloud.

The cold gas, which is as cool as the cold dust, 
constitutes approximately 25\% of the dense molecular gas 
based on the NH$_3$ emission.
Our complete survey of the CMZ shows that 
the dense molecular gas in the CMZ is dominated by 
gas that is warmer than the majority of dust present there.


\subsection{Comparison with Pressure Distribution}

Although we cannot estimate the molecular gas temperature in regions
where no NH$_3$ emission is detected,
we can make rough evaluations using other spectral lines.
Using the intensity ratios of 
$^{12}$CO ($J = 2\mbox{--}1$) over $^{12}$CO ($J = 1\mbox{--}0$), 
$R_{\mathrm{CO}(2\mbox{--}1)/(1\mbox{--}0)}$, 
and of $^{13}$CO ($J = 2\mbox{--}1$) over $^{12}$CO ($J = 2\mbox{--}1$), 
\citet{sawada} derived the gas pressure of molecular clouds in the Galactic 
center region.

Our NH$_3$ observations are made with the same beamsize and the same 
sampling grid as that in the observations in $^{12}$CO ($J = 1\mbox{--}0$) 
using the CfA 1.2-m telescope (\cite{bitran}) and 
$^{12}$CO ($J = 2\mbox{--}1$) using the Tokyo-Onsala-ESO-Cal\'an 60-cm 
telescope (\cite{sawada}).
Therefore, we can make direct comparisons with 
the $R_{\mathrm{CO}(2\mbox{--}1)/(1\mbox{--}0)}$ distribution.

\begin{figure*}[h]
  \begin{center}
    \FigureFile(160mm,80mm){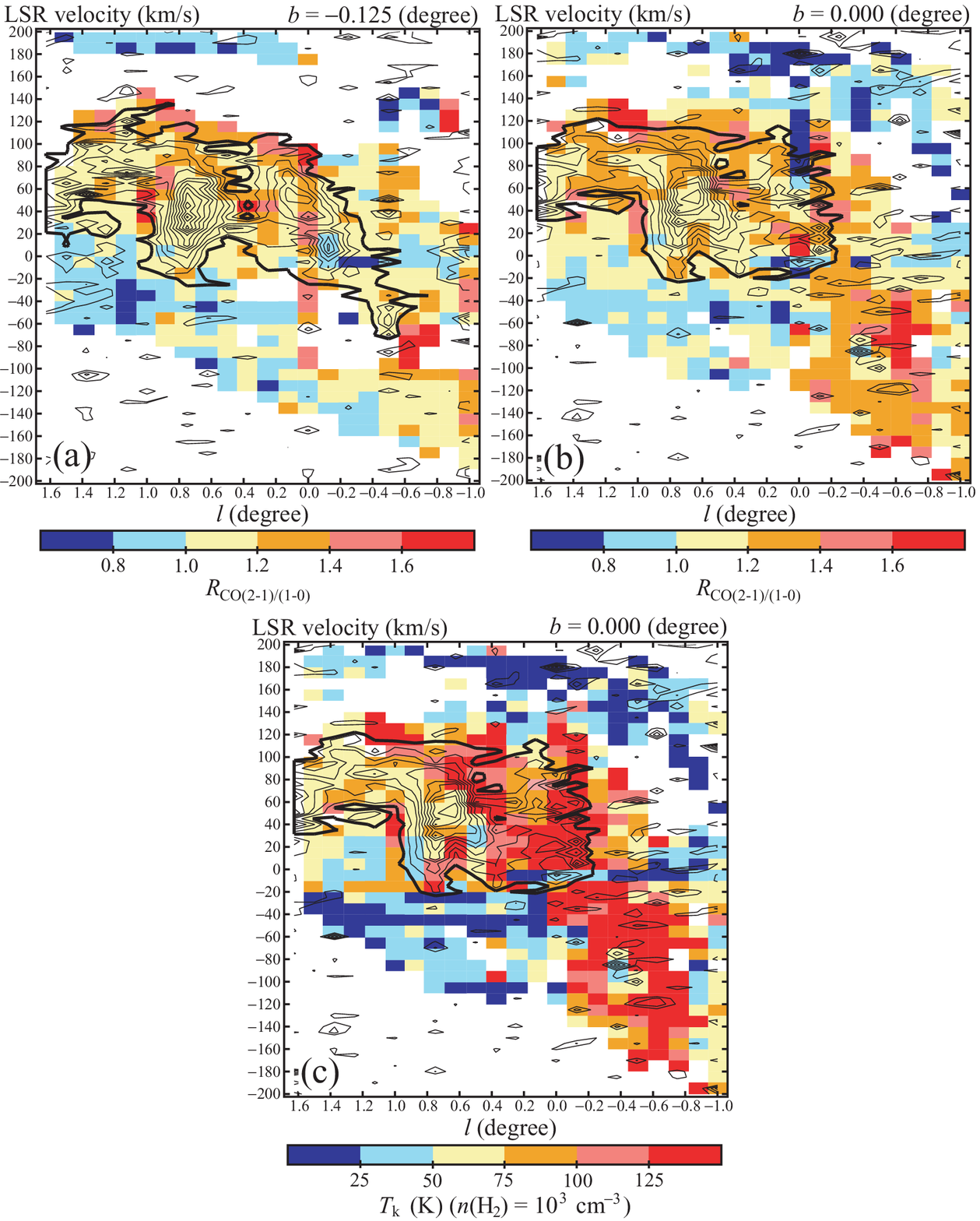}
    \caption{(a)(b): NH$_3$ (1,1) $l$-$v$ diagrams superimposed on
             the $R_{\mathrm{CO}(2\mbox{--}1)/(1\mbox{--}0)}$
             $l$-$v$ diagrams by \citet{sawada}.
             (c): NH$_3$ (1,1) $l$-$v$ diagram superimposed on
             the $T_{\mathrm{k}}$ $l$-$v$ diagram.
             $T_{\mathrm{k}}$ is derived from the LVG approximation for
             CO lines assuming $n(\mathrm{H}_2) = 10^{3}$ cm$^{-3}$.}
    \label{fig:co_ratio}
  \end{center}
\end{figure*}

Figure \ref{fig:co_ratio} shows an NH$_3$ (1,1) $l$-$v$ diagram 
superimposed on the $R_{\mathrm{CO}(2\mbox{--}1)/(1\mbox{--}0)}$ 
$l$-$v$ diagram by \citet{sawada}.
In the CO lines, the absorption caused by the foreground arms 
has an effect at $v_{\mathrm{LSR}} \sim -50, -30, 0$ km s$^{-1}$.
This absorption effect is not seen in NH$_3$ line.
When comparing the distributions of the NH$_3$ emission and 
$R_{\mathrm{CO}(2\mbox{--}1)/(1\mbox{--}0)}$ 
on the $l$-$v$ plane at $b = \timeform{-0D.125}$ and \timeform{0D.000}, 
the NH$_3$ emitting region is surrounded by 
the high $R_{\mathrm{CO}(2\mbox{--}1)/(1\mbox{--}0)}$ region.
In the Sgr B molecular cloud complex with its intense NH$_3$ emission,
$R_{\mathrm{CO}(2\mbox{--}1)/(1\mbox{--}0)}$ is not very high.

For $b = \timeform{0D.000}$, 
\citet{sawada} provide the gas pressure distribution using 
the large velocity gradient (LVG) approximation (\cite{gold}).
For $n(\mathrm{H}_2) = 10^{3}$ cm$^{-3}$, 
which is the NH$_3$ critical density (\cite{morris}; \cite{ho}),
we can also obtain the distribution of kinetic temperature from 
their LVG approximation
(see Figure \ref{fig:co_ratio} (c)).
This provides an upper limit for the temperature in the region
emitting NH$_3$,
and a lower limit for the region without NH$_3$ emission.

\begin{figure*}[h]
  \begin{center}
    \FigureFile(80mm,80mm){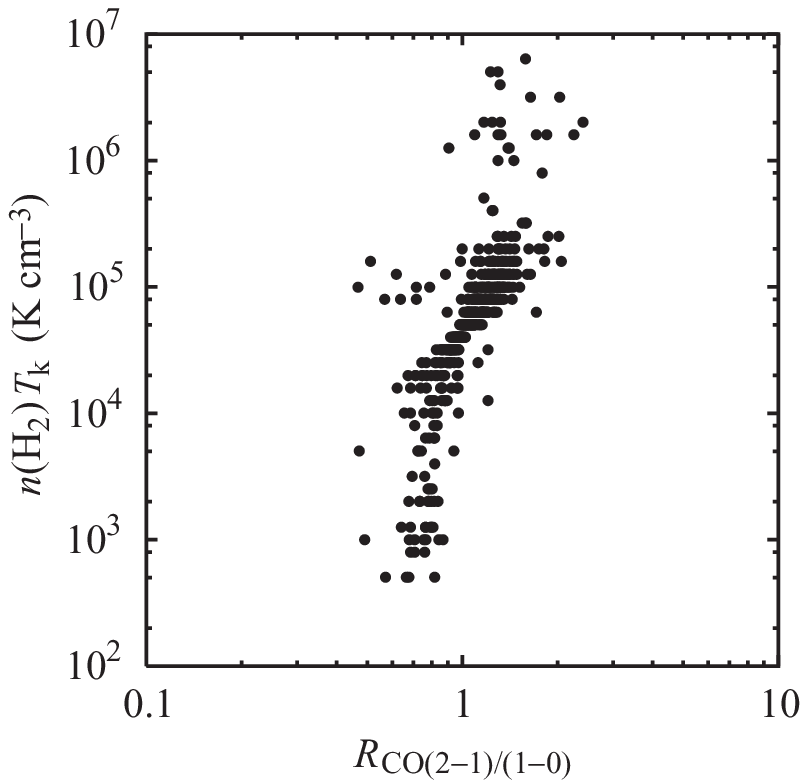}
    \caption{Correlation of $R_{\mathrm{CO}(2\mbox{--}1)/(1\mbox{--}0)}$
             and $n(\mathrm{H}_2) T_{\mathrm{k}}$.}
    \label{fig:relation}
  \end{center}
\end{figure*}

For $b = \timeform{-0D.125}$, 
we cannot obtain the $T_{\mathrm{k}}$ limit because \citet{sawada} derive 
$n$(H$_2) T_{\mathrm{k}}$ only for $b = \timeform{0D.000}$.
However, using the $b = \timeform{0D.000}$ data,
we find a correlation between $n$(H$_2$) $T_{\mathrm{k}}$ and 
$R_{\mathrm{CO}(2\mbox{--}1)/(1\mbox{--}0)}$, 
as shown in Figure \ref{fig:relation}.
The correlation is summarized as follows:
$n$(H$_2$) $T_{\mathrm{k}}$ increases with 
$R_{\mathrm{CO}(2\mbox{--}1)/(1\mbox{--}0)}$ and
$n$(H$_2$) $T_{\mathrm{k}} = 4 \times 10^{4}$ K cm$^{-3}$ when 
$R_{\mathrm{CO}(2\mbox{--}1)/(1\mbox{--}0)}$ = 1.0.
This implies that the high $R_{\mathrm{CO}(2\mbox{--}1)/(1\mbox{--}0)}$ value 
indicates a high pressure region that can in principle be dense or hot.

Although NH$_3$ emission can trace dense gas, 
it does not extend into the high pressure region, 
at least at high intensities.
Therefore, the high pressure region is less dense and hotter.
In the Galactic center, the molecular cloud complex has 
``cores'' of dense and warm clouds which are traced in the NH$_3$ emission, 
and an ``envelope'' of less dense and hotter gas clouds.
For the cores, the pressure is found to be
$n(\mathrm{H}_2) T_{\mathrm{k}} \sim 5\mbox{--}10 \times 10^{4}$ K cm$^{-3}$ 
from the LVG approximation and 
the kinetic temperature is $T_{\mathrm{k}} \sim$ 20--100 K based on 
our NH$_3$ ratio.
In the case where the molecular gas in a beam is in pressure equilibrium,
the density is $n(\mathrm{H}_2) \gtrsim 10^{3}$ cm$^{-3}$.
For the envelope, the pressure is found to be
$n(\mathrm{H}_2) T_{\mathrm{k}} \geq 10^{5}$ K cm$^{-3}$ from 
the LVG approximation.
The density $n(\mathrm{H}_2$) should be less than
$10^{3}$ cm$^{-3}$ in the envelope because of the absence of NH$_3$ emission 
there.
This reveals that the kinetic temperature of the envelope is at least 100 K.


\subsection{Origin of the High Pressure Region}

In the previous section, we showed that the large scale structure 
of the molecular cloud complex in the CMZ exhibits a dense core 
surrounded by a hot envelope.
What is the heating source of the envelope?
There are four possibilities.

\begin{figure*}[h]
  \begin{center}
    \FigureFile(160mm,80mm){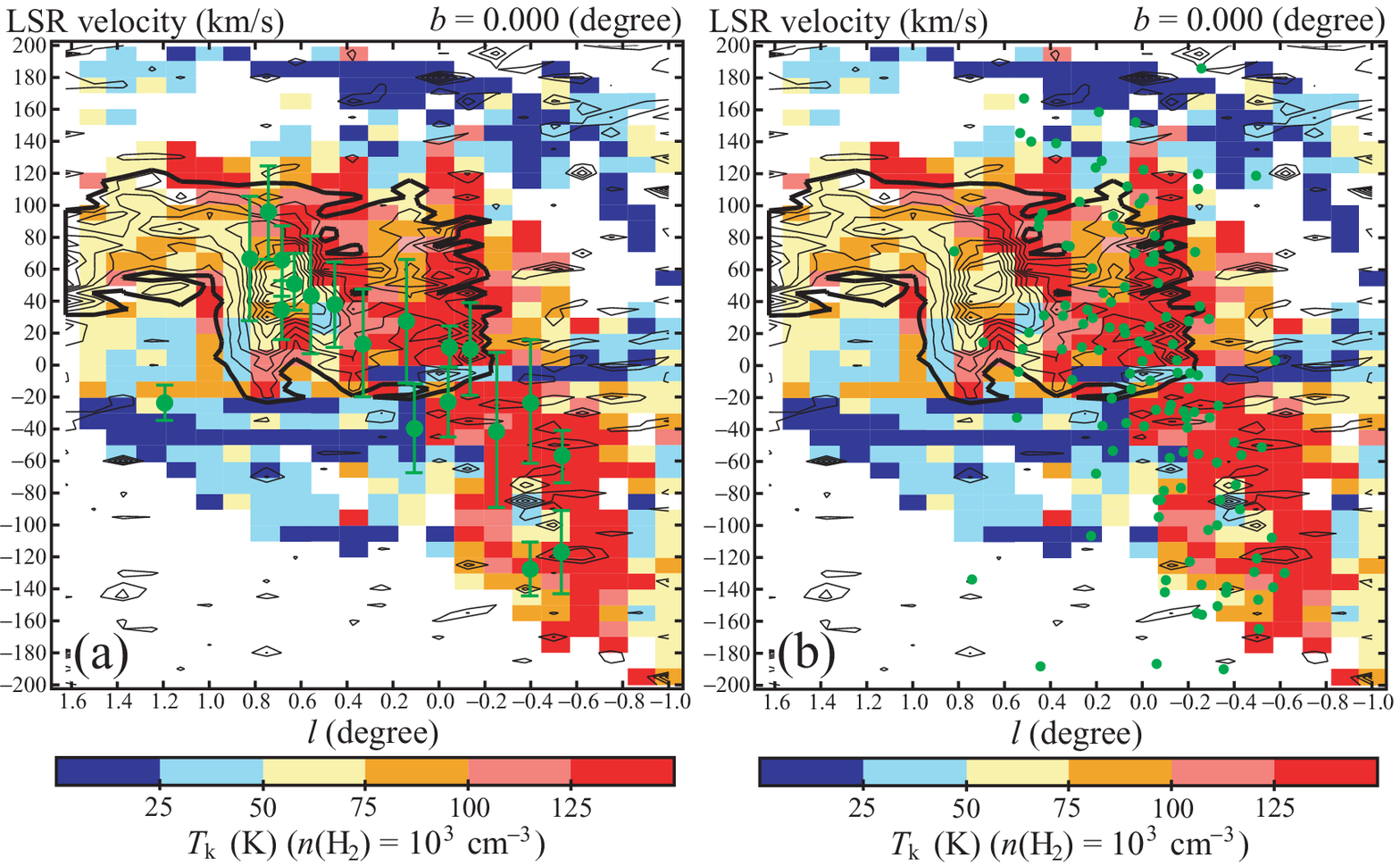}
    \caption{(a): The $l$-$v$ distribution of the H109$\alpha$ hydrogen
             recombination line at 5 GHz (\cite{pauls}) superimposed
             on the NH$_3$ (1,1) and the $T_{\mathrm{k}}$ $l$-$v$ diagrams.
             (b): The $l$-$v$ distribution of OH/IR stars observed
             in the VLA survey (\cite{lind}) superimposed
             on the NH$_3$ (1,1) and the $T_{\mathrm{k}}$ $l$-$v$ diagrams.
             The OH/IR star survey area is
             \timeform{-1D} $\leq l \leq$ \timeform{1D}.}
    \label{fig:ohir}
  \end{center}
\end{figure*}

The first is radiation from massive stars.
For $l<$ \timeform{0D.35},
the H\emissiontype{II} regions traced by the H109$\alpha$ line (\cite{pauls})
appear in the high pressure region (see Figure \ref{fig:ohir} (a)).
This suggests that the high pressure region is heated by 
intense radiation from the massive OB stars.

This model appears to be invalid for $l>$\timeform{0D.35}, 
because the H\emissiontype{II} regions appear 
in the core of the Sgr B cloud that exhibits intense NH$_3$ emission.
Therefore, a hot gas heated by intense radiation should be present there.
However, we believe this feature is due to a dilution effect. 
The hot gas must be compact and surrounded by a large amount of cold gas.
In this case, 
the derived gas temperature with our \timeform{9'.5} beam should be low.
In fact, hot and compact gas regions in Sgr B are detected 
by high resolution observations.
\citet{hutt2} obtained two emission cores of \timeform{4"} at
$v_{\mathrm{LSR}} = 65$ km s$^{-1}$ with $T_{\mathrm{k}} =$ 150--300 K 
in Sgr B2.
They also found the 85 km s$^{-1}$ component with $T_{\mathrm{k}} = 160$ K 
in Sgr B2(N).
Actually, the 85 km s$^{-1}$ component, 
which is as large as \timeform{120"}, 
is detected in the high pressure region.

There is further evidence of star formation during a short period 
in the high pressure region.
In the high pressure region, 134 OH/IR stars observed using VLA 
(\cite{lind}) are located (see Figure \ref{fig:ohir} (b)).
This suggests that star formation was active during a short period 
in the high pressure region.
These stars should heat the gas by radiation during this period.
Although intense radiation from the OB stars heats the molecular gas,
late OH/IR stars are also sufficient to heat the dust and gas.
However, the dust is known to be cool.
Photons with energy below 13.6 eV, which cannot ionize the hydrogen, 
can heat warm gas, thereby transforming it to hot gas.
Using the blackbody spectra and the 13.6 eV cutoff, 
we can estimate the equivalent number of OH/IR stars required for 
an OB star to heat up the molecular gas.
The typical temperatures of an OH/IR star and an OB star are
3000 K and 30000 K, respectively.
The typical absolute magnitude of an OB star is $M_V = -6$.
For an OH/IR star, the typical absolute magnitude is $M_V = -4$ 
and $M_V = -6$ in the case of a supergiant.
Using these values, 
we find that approximately 10 OH/IR stars or 1 OH/IR supergiant
is equivalent to an OB star.
This implies that OH/IR and OB stars are comparable heating sources.
The distribution of OH/IR stars on the $l$-$v$ plane suggests that
they heat the molecular gas in the high pressure region.

However, stellar radiation may not be a major heating mechanism for 
the CMZ gas.
The stellar radiation heats the molecular gas through dust heating.
In that case, the dust temperature should be higher than the gas temperature.
However, our observations indicate that the major portion of 
the molecular gas is warmer than the dust.

Heating by extremely hot plasma gas emitting X-ray is the second possibility.
X-ray emission and radio continuum are intense for $l<$\timeform{0D.35}
(\cite{wang}; \cite{handa}).
This is consistent with the high pressure region, 
which is conspicuous for $l<$\timeform{0D.35}.

The cooling luminosity of clouds in the CMZ exceeds 
the X-ray luminosity by a factor of 40--400 (\cite{morris}).
Therefore, it seems that the gas cannot be heated by X-ray emission itself.
However, the distribution of the hot plasma gas emitting X-ray
is well coincident with the high pressure region.
The total mass of the hot molecular gas in the high pressure region is 
$2 \times 10^{6} \MO$ 
from the CO line intensity with the conversion factor
$X = 2 \times 10^{19}$ cm$^{-2}$ (K km s$^{-1} )^{-1}$ (\cite{dahmen}).
The total thermal energy required to heat the warm gas (30 K) to 
the hot gas (100 K) is $3 \times 10^{49}$ erg.
Conversely, in X-rays, the total thermal energy radiated from 
the hot plasma gas with a temperature of $10^{8}$ K is 
(4--8) $\times 10^{53}$ erg (\cite{yamauchi}).
This value is $10^{4}$ times larger than the energy required to 
heat the warm gas to hot gas.
Therefore, the hot plasma gas emitting X-ray can heat the hot gas 
in the high pressure region with any in sufficient processes.

The third is 
the shock heating through cloud-cloud collisions.
This mechanism is suggested by \citet{gusten2} and \citet{hutt}.
In the case of the cloud-cloud collisions, 
the gas temperature increases with 
the density and the line width (\cite{gusten2}).
However, our result shows the high pressure region is less dense 
than the NH$_3$ emitting region.
\citet{hutt} reported that
there is no good correlation between the temperature and the line width.
These results suggest that the cloud-cloud collisions heating
is not strongly supported.
However, the energy of the turbulent motions 
in the clouds within 300 pc of the center is larger than
the thermal energy required to heat to the hot gas, $10^{49}$ erg.
The energy of the turbulent motions is $10^{53}$ erg (\cite{gusten2}).
The heating through cloud-cloud collisions cannot be ruled out by our data.

The last is the heating of the ambipolar diffusion.
\citet{hutt} suggests that the gas can bring to $\sim$ 200 K,
if the ionization fraction is of order $10^{-8}$ and
the magnetic field strength is 500 $\mu$G.


\subsection{Star Formation in the Galactic Center}

In the CMZ, the distribution of H\emissiontype{II} regions extends
more toward the Galactic eastern side than the OH/IR stars.
This distribution difference suggests that
star formation activity moves from the Galactic west to Galactic east.
Stars become OH/IR stars $10^{8}$ years after their formation (\cite{oka1}).
The distribution shows that in the high pressure region, 
star formation was active $10^{8}$ years ago.
The lifetime of an H\emissiontype{II} region is 
$10^{5}$ years and OB stars are younger than $10^{6}$ years.
Therefore, the timescale of the transition of star formation is $10^{8}$ 
years.


\subsection{Physical Conditions of the Molecular Clouds}

\begin{figure*}[h]
  \begin{center}
    \FigureFile(160mm,40mm){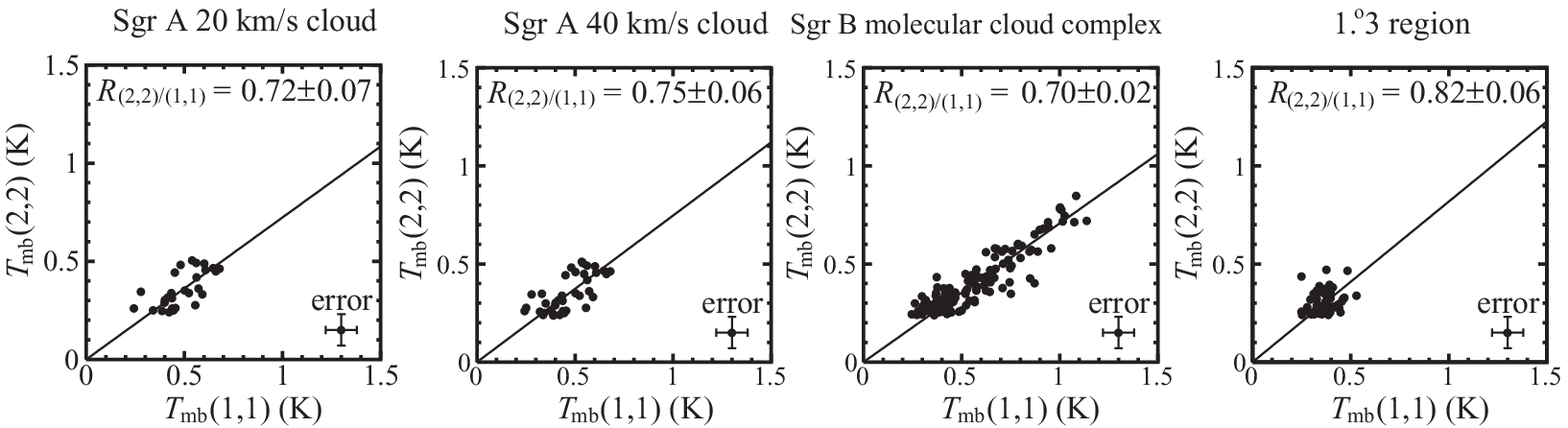}
  \end{center}
  \caption{Correlations of the main beam temperature of the NH$_{3}$
           (1,1) and (2,2) lines for four clouds, respectively.}
  \label{fig:1vs2}
\end{figure*}

For the four major clouds, 
we estimate the average intensity ratio of the NH$_3$ (2,2) to (1,1) lines
using a correlation plot between these two lines.
The estimated ratio and 
the derived rotational and kinetic temperatures
are shown in Figure \ref{fig:1vs2} and Table \ref{tab:kinetic}.
In order to reduce the noise fluctuation,
we only use those pixels where
the (1,1) and (2,2) intensities exceed the 3 $\sigma$ level.
All the four clouds show almost the same ratio of approximately 0.75.
The rotational temperature corresponding to the observed ratio of 
0.7--0.8 is 36--42 K in the optically thin case and 
24--34 K the in optically thick case.
This implies that the kinetic temperatures of these Galactic center clouds
are higher than those of the Galactic disk clouds.
The temperature of 40 massive protostar candidates is 15--20 K 
(\cite{srid}) and 
that of quiescent cores near the Ori GMC is 14--19 K (\cite{li}).

Table \ref{tab:condition} lists the physical parameters of 
each of the four clouds estimated under the optically thin ($\tau \ll 1$) 
assumption.
The sizes of four clouds are determined from the FWHM 
of the contour in the $l$-$v$ diagram (see Figure \ref{fig:schematic}).
The estimated column densities for the two metastable levels
from the integrated line intensities are
$N(1,1) = $(2.5--11) $\times 10^{14}$ cm$^{-2}$ and 
$N(2,2) = $(1.2--5.5)$\times 10^{14}$ cm$^{-2}$.
Using the abundance ratio of $X(\mathrm{NH}_3) =10^{-9}$ (\cite{hutt}) and
spherically symmetric geometry,
the H$_2$ number density is $n$(H$_2$) $\sim 10^{3}$--$10^{4}$ cm$^{-3}$.
The luminosity mass, which is estimated from the integration of the column 
densities, is $M_{\mathrm{lum}} = (2.7\mbox{--}23) \times 10^{6} \MO$.
Using the method of \citet{mcgary},
the estimated intrinsic line width is approximately 5--10 km s$^{-1}$ 
smaller than the observed line width in the observed line width range.
Using the derived intrinsic line width, 
the virial mass is estimated to be 
$M_{\mathrm{vir}} = (2.8\mbox{--}30) \times 10^{6} \MO$.
Among these four clouds, the Sgr B molecular cloud complex 
is the densest and most massive.

The luminosity mass and the virial mass are consistent 
in the order of magnitude.
This suggests that the optically thin assumption is appropriate 
for these clouds. 


\subsection{Origin of the \timeform{0D.9} Wing Feature}

At $l = \timeform{0D.9}$ we find a unique feature (see section 3.6).
A blueshifted wing as wide as 50 km s$^{-1}$ is observed.
The estimated ratio $R_{(2,2)/(1,1)}$ of the 20 km s$^{-1}$ component, 
which comprises the \timeform{0D.9} wing feature, is $0.89 \pm 0.12$ 
(see Table \ref{tab:l0.8}).
To derive $R_{(2,2)/(1,1)}$,
the components near $l = \timeform{0D.9}$ are integrated. 
The integrated velocity ranges of the 20 km s$^{-1}$ and 
80 km s$^{-1}$ components are $-30 < v_{\mathrm{LSR}} < 50$ km s$^{-1}$
and $50 < v_{\mathrm{LSR}} < 120$ km s$^{-1}$, respectively.
The estimated ratio is higher than the ratios of other components:
$0.69 \pm 0.08$ for the 80 km s$^{-1}$ component and
0.7--0.8 for the four identified clouds shown in section 4.5.
The rotational temperature corresponding to 
$R_{(2,2)/(1,1)} = 0.89 \pm 0.12$
is $45\pm6$ K in the optically thin case and 
$42^{+17}_{-13}$ K in the optically thick case.

This high ratio is also seen in previous observations.
This ratio is consistent with the value shown in Fig. 4 of \citet{morris},
who, however did not mention this high ratio feature.
A part of the \timeform{0D.9} wing feature is 
identified as M$+0.83-0.10$ by \citet{hutt}.
The $R_{(2,2)/(1,1)}$ for M$+0.83-0.10$ has a higher value, 
$R_{(2,2)/(1,1)} = 1.1$.

The size of this feature is estimated to be approximately one beamsize 
\timeform{9'.5} = 24 pc
because evidence of the feature is seen 
in two profiles spaced \timeform{0D.125} = 18 pc apart.
The mass $M_{\mathrm{lum}} = 1.8 \times 10^{6} \MO$ 
is derived from the column density, 
$N(\mathrm{NH}_3) = 4.2 \times 10^{14}$ cm$^{-2}$,
by assuming the abundance ratio, $X(\mathrm{NH}_3) = 10^{-9}$,
and optically thin emission. 
The velocity offset from the 20 km s$^{-1}$ component 
is $\Delta v = 50$ km s$^{-1}$.
The kinetic energy is $10^{52}$ erg in the case where
the 20 km s$^{-1}$ component is ejected from 
the 80 km s$^{-1}$ component.

What is the origin of this feature?
There are two possibilities.

The morphology in the $l$-$v$ diagram and profiles suggest that
the \timeform{0D.9} wing feature may be ejected from the Sgr B cloud.
It should have been ejected by star formation activity and/or 
supernova (SN) explosions.
The higher gas temperature supports this idea.
However, the kinetic energy of the component requires 
$10^3$ SN explosions.
Therefore, we do not favor the explosion approach to explain the origin.
In the case of an explosive origin, 
a tracer of lower density gas should exhibit a wider wing there.
However, we cannot confirm this feature, 
because the wing is overlapped by foreground gas in the spiral arms.

The other possibility is gas infall from the off-plane.
There are several models that suggest the exchange of gas 
between the disk and halo.
Some of them show that infalling gas is accumulated by
the Galactic magnetic field (\cite{matsu}).
The \timeform{0D.9} wing feature may be a compact and in-plane version 
of the molecular spur found in NGC 891 (\cite{handa2}).

In the NRO CS survey, several high velocity wing features 
like the \timeform{0D.9} wing feature at other positions
were observed (\cite{tsuboi}).
None of them, except the \timeform{0D.9} wing feature,
exhibit any wings in our survey.
However, most of them may exhibit a similar wing feature in 
high resolution observations of NH$_3$.


\section{Conclusions}

We have presented a map of the major part of the CMZ 
by simultaneous observations in NH$_3$ $(J,K)$ = (1,1) and (2,2) lines 
with the Kagoshima 6-m telescope. 
Given below are our observations and conclusions:

\begin{enumerate}

  \item 
From the $l$-$b$-$v$ data cube,
we identified and investigated four clouds, 
which correspond to Sgr A 20 km s$^{-1}$ cloud, 
Sgr A 40 km s$^{-1}$ cloud, 
Sgr B molecular cloud complex, and the \timeform{1D.3} region.

  \item 
We found a unique ``\timeform{0D.9} wing feature'' which 
is a prominent high velocity wing.
$R_{(2,2)/(1,1)}$ has a slightly higher value of $0.89 \pm 0.12$.
In an ejection scenario, 
the kinetic energy is estimated to be $10^{52}$ erg, 
which requires $10^3$ SN explosions.
The other possibility of its origin is infall from off-plane.

  \item 
The kinetic temperatures derived from the (2,2) to (1,1) intensity 
ratios are 20--80 K or exceed 80 K.
The gases corresponding to temperature of 20--80 K and $\geq$ 80 K 
contain 75\% and 25\% of the total NH$_3$ flux, respectively.
Our complete survey of the CMZ showed that 
the dense molecular gas in the CMZ is dominated by gas
that is warmer than the majority of dust present in that region.

  \item
A comparison with the CO survey by \citet{sawada}
showed the molecular cloud complex in the Galactic center has
a core of dense and warm clouds, 
and an envelope of less dense and hotter gas clouds.
The heating mechanisms of the hot envelope are 
still in mystery, we propose heating by
the thermal energy of the hot plasma gas observed in X-ray emission,
besides by ambipolar diffusion.

  \item
H\emissiontype{II} regions are distributed over 
the Galactic east side to a greater extent than OH/IR stars.
This distribution difference shows
the transition of star formation from the Galactic west to Galactic east.

  \item
We present the physical conditions of the four clouds
estimated under the optically thin assumption.
All the clouds have an almost similar $R_{(2,2)/(1,1)}$ value of 
approximately 0.75.
The kinetic temperatures of these Galactic center clouds
are higher than those of the Galactic disk clouds.

\end{enumerate}


\bigskip

We thank K. Takeda, a student of the Kagoshima University, 
for his support in the observations. 
We also thank N. Matsuyama, A. Hasegawa, and S. Morisaki,
who are Kagoshima University graduates,
for their support in the observations.
We acknowledge K. Miyazawa for his technical support and 
T. Hirota and H. Kobayashi for their scientific advices.
We thank the referee for suggestions that improved the paper.
T.O. was supported by a Grant-in-Aid for Scientific Research 
from the Japan Society for Promotion Science (17340055).
T.H. thanks the Japan Society for Promotion of 
Science for the financial support of provided by 
the JSPS Grant-in-Aid for Scientific Research C (18540232) and 
S (17104002).


\begin{table*}[h]
\begin{center}
\caption{The integrated intensity distribution
         of the NH$_{3}$ (1,1) and (2,2) lines.}
\label{tab:flux}
 \begin{tabular}{lrrrrrr}
  \hline \hline
  \multicolumn{1}{c}{Feature}                       &
  \multicolumn{2}{c}{$\int T_{\mathrm{mb}}(1,1) dv$} &
  \multicolumn{2}{c}{$\int T_{\mathrm{mb}}(2,2) dv$} &
  \multicolumn{2}{c}{$\int T_{\mathrm{mb}}($NH$_{3}) dv$} \\
                                                    &
  \multicolumn{2}{c}{(K km s$^{-1}$)}               &
  \multicolumn{2}{c}{(K km s$^{-1}$)}               &
  \multicolumn{2}{c}{(K km s$^{-1}$)}               \\
  \hline
  The whole observed area                         & 1580&      & 1330&      & 2910&       \\
  Galactic eastern side ($l \geq $ \timeform{0D}) & 1290&(82\%)& 1080&(81\%)& 2370&(81\%) \\
  Galactic western side ($l <$ \timeform{0D})     &  290&(18\%)&  250&(19\%)&  540&(19\%) \\
  Galactic northern side ($b \geq $\timeform{0D}) &  690&(44\%)&  580&(44\%)& 1270&(44\%) \\
  Galactic southern side ($b <$ \timeform{0D})    &  890&(56\%)&  750&(56\%)& 1640&(56\%) \\
  Galactic plane ($b = \timeform{0D}$)            &  460&(29\%)&  330&(25\%)&  790&(27\%) \\
  Sgr A                                           &  300&(19\%)&  250&(19\%)&  550&(19\%) \\
  Sgr B                                           &  550&(35\%)&  410&(31\%)&  960&(33\%) \\
  The \timeform{1D.3} region                      &  160&(10\%)&  130&(10\%)&  290&(10\%) \\
  \hline
  \end{tabular}
\end{center}
\end{table*}

\begin{table*}
\begin{center}
\caption{Conversion from the ratio of NH$_3$ (2,2) to (1,1) lines to
         the rotational and kinetic temperatures.
         The rotational temperatures are estimated under
         the optically thin ($\tau \ll 1$) and
         optically thick ($\tau \sim 10$) assumption
         using the method of \citet{morris}.
         The conversion of the rotational temperature to
         kinetic temperature is based on \citet{hutt}.}
\label{tab:2}
    \begin{tabular}{rrrrr}
    \hline \hline
        \multicolumn{1}{c}{$R_{(2,2)/(1,1)}$}     &
        \multicolumn{2}{c}{$T_{\mathrm{rot}}$ (K)} &
        \multicolumn{2}{c}{$T_{\mathrm{k}}$ (K)}   \\
                                           &
        \multicolumn{1}{c}{$\tau \ll 1$}   &
        \multicolumn{1}{c}{$\tau \sim 10$} &
        \multicolumn{1}{c}{$\tau \ll 1$}   &
        \multicolumn{1}{c}{$\tau \sim 10$} \\
    \hline
        0.5     & 27    & 17    & 35            & 17 \\
        0.6     & 32    & 20    & 45            & 20 \\
        0.7     & 36    & 25    & 60            & 30 \\
        0.8     & 41    & 32    & $>$ 80        & 45 \\
        0.9     & 46    & 43    & $>$ 80        & $> $80 \\
        1.0     & 52    & 67    & $>$ 80        & $> $80 \\
    \hline
    \end{tabular}
\end{center}
\end{table*}

\begin{table*}[h]
\begin{center}
\caption{Estimated average intensity ratio of the NH$_3$ (2,2) to (1,1) lines,
         rotational temperatures, and kinetic temperatures
         of the four clouds.
         The rotational temperatures and the kinetic temperatures
         are derived in the optically thin ($\tau \ll 1$) and
         optically thick ($\tau \sim 10$) cases.}
\label{tab:kinetic}
    \begin{tabular}{clccccc}
    \hline \hline
        ID                                        &
        \multicolumn{1}{c}{Feature}               &
        $R_{(2,2)/(1,1)}$                         &
        \multicolumn{2}{c}{$T_{\mathrm{rot}}$ (K)} &
        \multicolumn{2}{c}{$T_{\mathrm{k}}$ (K)}   \\
                       &
                       &
                       &
        $\tau \ll 1$   &
        $\tau \sim 10$ &
        $\tau \ll 1$   &
        $\tau \sim 10$ \\
    \hline
        1 ...... & Sgr A 20 km s$^{-1}$ cloud     & $0.72\pm0.07$ & $37\pm3 $ & $26\pm4$ & $64\pm7 $ & $35\pm8$  \\
        2 ...... & Sgr A 40 km s$^{-1}$ cloud     & $0.75\pm0.06$ & $38\pm3 $ & $28\pm4$ & $68\pm7 $ & $40\pm8$  \\
        3 ...... & Sgr B molecular cloud complex  & $0.70\pm0.02$ & $36\pm1 $ & $24\pm1$ & $60\pm2 $ & $32\pm2$  \\
        4 ...... & \timeform{1D.3} region         & $0.82\pm0.06$ & $42\pm3 $ & $34\pm6$ & $> 80   $ & $56\pm16$ \\
    \hline
    \end{tabular}
\end{center}
\end{table*}

\begin{table*}[h]
\begin{center}
\caption{Derived parameters of the four clouds.}
\label{tab:condition}
    \begin{tabular}{clrrrrrrrr}
    \hline \hline
        ID                                    &
        \multicolumn{1}{c}{Feature}           &
        \multicolumn{1}{c}{$S$}               &
        \multicolumn{1}{c}{$N(1,1)$}          &
        \multicolumn{1}{c}{$N(2,2)$}          &
        \multicolumn{1}{c}{$N$(NH$_3$)}       &
        \multicolumn{1}{c}{$n$(NH$_3$)}       &
        \multicolumn{1}{c}{$n$(H$_2$)}        &
        \multicolumn{1}{c}{$M_{\mathrm{vir}}$} &
        \multicolumn{1}{c}{$M_{\mathrm{lum}}$} \\
                                        &
                                        &
        \multicolumn{1}{c}{(pc)}        &
        \multicolumn{1}{c}{(cm$^{-2}$)} &
        \multicolumn{1}{c}{(cm$^{-2}$)} &
        \multicolumn{1}{c}{(cm$^{-2}$)} &
        \multicolumn{1}{c}{(cm$^{-3}$)} &
        \multicolumn{1}{c}{(cm$^{-3}$)} &
        \multicolumn{1}{c}{(\MO)}       &
        \multicolumn{1}{c}{(\MO)}       \\
                                        &
                                        &
                                        &
        \multicolumn{1}{c}{$10^{14}$}   &
        \multicolumn{1}{c}{$10^{14}$}   &
        \multicolumn{1}{c}{$10^{14}$}   &
        \multicolumn{1}{c}{$10^{-6}$}   &
        \multicolumn{1}{c}{$10^{3}$}    &
        \multicolumn{1}{c}{$10^{6}$}    &
        \multicolumn{1}{c}{$10^{6}$}    \\
    \hline
      1 ...... & Sgr A 20 km s$^{-1}$   & 20 & 4.5 & 2.1 & 6.6 & 11  & 11  & 2.8  & 2.7 \\
      2 ...... & Sgr A 40 km s$^{-1}$   & 33 & 3.7 & 2.1 & 5.9 & 5.7 & 5.7 & 12   & 8.3 \\
      3 ...... & Sgr B                  & 33 & 11  & 5.5 & 16  & 16  & 16  & 30   & 23  \\
      4 ...... & \timeform{1D.3} region & 28 & 2.5 & 1.2 & 3.8 & 4.2 & 4.2 & --\footnotemark[$*$] & 3.9 \\
    \hline
    \multicolumn{10}{@{}l@{}} {\hbox to 0pt{\parbox{70mm} {\footnotesize
        \footnotemark[$*$]
                The virial mass of \timeform{1D.3} region is not derived,
                because signal to noise ratio is too poor to estimate
                an accurate line width.
        \par\noindent
        }\hss}}
    \end{tabular}
  \end{center}
\end{table*}

\begin{table*}[h]
\begin{center}
\caption{$R_{(2,2)/(1,1)}$ of the components near $l = \timeform{0D.9}$.}
\label{tab:l0.8}
 \begin{tabular}{rrrr}
  \hline \hline
        \multicolumn{1}{c}{$l$}   &
        \multicolumn{1}{c}{$b$}   &
        \multicolumn{2}{c}{$R_{(2,2)/(1,1)}$} \\
        \multicolumn{1}{c}{(deg)} &
        \multicolumn{1}{c}{(deg)} &
        \multicolumn{1}{c}{20 km s$^{-1}$\footnotemark[$*$]} &
        \multicolumn{1}{c}{80 km s$^{-1}$\footnotemark[$\dagger$]} \\
  \hline
  $0.875$\footnotemark[$\ddagger$] & $-0.125$       & $0.82 \pm 0.11$ & $0.64 \pm 0.09$ \\
  $0.875$\footnotemark[$\ddagger$] & $ 0.000$       & $0.71 \pm 0.07$ & $0.70 \pm 0.04$ \\
  $0.880$\footnotemark[$\S$]       & $ 0.000$       & $1.15 \pm 0.19$ & $0.73 \pm 0.10$ \\
  \hline
  \multicolumn{2}{c}{Mean}         & $0.89 \pm 0.12$ & $0.69 \pm 0.08$ \\
  \hline
  \multicolumn{4}{@{}l@{}} {\hbox to 0pt{\parbox{70mm} {\footnotesize
        \footnotemark[$*$]
                The 20 km s$^{-1}$ component.
                Velocity range is $-30 < v_{\mathrm{LSR}} < 50$ km s$^{-1}$.
        \par\noindent
        \footnotemark[$\dagger$]
                The 80 km s$^{-1}$ component.
                Velocity range is $50 < v_{\mathrm{LSR}} < 120$ km s$^{-1}$.
        \par\noindent
        \footnotemark[$\ddagger$]
                Observed with the Kagoshima 6-m telescope.
        \par\noindent
        \footnotemark[$\S$]
                Observed with the Kashima 34-m telescope.
        \par\noindent
        }\hss}}
  \end{tabular}
\end{center}
\end{table*}



\end{document}